\documentclass[conference]{IEEEtran}
\IEEEoverridecommandlockouts
\usepackage{cite}
\usepackage{amsmath,amssymb,amsfonts}
\usepackage{graphicx}
\usepackage{textcomp}
\usepackage{xcolor}
\usepackage{booktabs}
\usepackage{marvosym}
\graphicspath{{figures/}}
\def\BibTeX{{\rm B\kern-.05em{\sc i\kern-.025em b}\kern-.08em
    T\kern-.1667em\lower.7ex\hbox{E}\kern-.125emX}}

\newcommand{\tool}{\textsc{MetaMorphQ}}

\begin{document}

\title{MetaMorphQ: Physics-Based Metamorphic Testing of Variational Quantum Circuits}

\author{%
\IEEEauthorblockN{%
\begin{tabular}{c@{\hspace{3em}}c@{\hspace{3em}}c}
Ngoc Nhi Nguyen\textsuperscript{\Letter} & John Le & Thai T. Vu\textsuperscript{\Letter} \\
\textit{University of Wollongong} & \textit{University of Wollongong} & \textit{University of Wollongong} \\
Australia & Australia & Australia \\
nnn469@uowmail.edu.au & johnle@uow.edu.au & tienv@uow.edu.au \\
0009-0006-8228-3615 & 0000-0003-0019-0345 & 0000-0002-9826-3321
\end{tabular}\\[1.5em]
\begin{tabular}{c@{\hspace{3em}}c}
Thi Thuy Nga Nguyen & Jun Shen \\
\textit{University of Wollongong} & \textit{University of Wollongong} \\
Australia & Australia \\
ttnn278@uowmail.edu.au & jshen@uow.edu.au \\
0009-0008-4777-5002 & 0000-0002-9403-7140
\end{tabular}%
}
}

\maketitle

\begin{abstract}
Variational Quantum Eigensolvers (VQEs) are central to quantum computing, yet testing them remains challenging due to the oracle problem: the ground-state energy they compute is itself unknown. Existing approaches, such as convergence-based testing, are unreliable and yield high false-positive rates due to optimisation instability. We propose METAMORPHQ, a metamorphic testing framework that derives test oracles directly from quantum mechanical properties of VQE circuits. Exploiting algebraic properties of parametrised rotation gates and diagonal Hamiltonians, we define five physics-based invariants that hold for any correct circuit and can be verified at initialisation without ground-truth outputs. Evaluated on 500 benchmark circuits with 2,469 mutants, METAMORPHQ achieves zero false positives and significantly improves diagnostic effectiveness (Youden’s J = 0.57 vs. 0.02 for convergence testing). These results demonstrate that physics-derived invariants provide a practical, oracle-free foundation for testing quantum software, enabling reliable validation of both human- and LLM-generated circuits.
\end{abstract}

\begin{IEEEkeywords}
metamorphic testing, variational quantum eigensolver, quantum software testing, mutation testing, physics-based invariants
\end{IEEEkeywords}

\section{Introduction}\label{sec:intro}

Variational Quantum Algorithms (VQAs) are among the leading candidates for practical computation on today's noisy intermediate-scale quantum (NISQ) devices~\cite{preskill2018nisq}. The Variational Quantum Eigensolver ~\cite{peruzzo2014vqe,mcclean2016theory} is the most prominent member of this family: it encodes a computational task as a quantum Hamiltonian and trains a parametrised circuit (the ansatz) to approximate its lowest-energy eigenstate via a hybrid quantum-classical optimisation loop. VQE has found use in quantum chemistry~\cite{kandala2017hardware}, combinatorial optimisation through QAOA~\cite{farhi2014qaoa}, and a range of further applications~\cite{tilly2022vqe_review}. As these implementations grow in complexity, ensuring their correctness becomes critical. VQE circuits are susceptible to subtle programming faults, such as swapping one rotation gate for another ($R_z \leftrightarrow R_y$), placing a gate on the wrong qubit, or corrupting a parameter, that produce no runtime error but silently alter the computed energy landscape.

Detecting such faults requires comparing the VQE output against a known correct value, but two fundamental obstacles make this impractical. The first is the \emph{oracle problem}: the ground-state energy $E_0$ is exactly what the VQE is trying to compute, so there is no independent reference to compare against~\cite{chen2018metamorphic,miranskyy2021testing}. For small instances ($n \leq 16$ qubits), $E_0$ can be found by classical diagonalisation, but the $O(2^n)$ cost makes this infeasible at the scales where quantum algorithms provide an advantage. The second is \emph{unreliable convergence}: the standard fallback is to run the VQE and check whether the energy converges to the expected value, but barren plateaus~\cite{mcclean2018barren,cerezo2021barren} and sensitivity to initialisation~\cite{grant2019initialization} prevent many correct circuits from converging within a practical budget. In our experiments on 500 circuits, 58.8\% of correct circuits do not converge within 400 steps. These two problems compound: without an oracle, the only fallback is convergence, but convergence itself is unreliable.

Several quantum software testing methods address parts of this challenge. Coverage-guided approaches~\cite{wang2021quito,ali2021assessing} systematically generate test inputs; differential testing~\cite{wang2022qdiff} catches inconsistencies across backends; statistical assertions~\cite{huang2019statistical} verify distributional patterns; property-based testing~\cite{honarvar2020property} checks universal properties like unitarity; and mutation testing frameworks~\cite{fortunato2022mutation,mendiluze2022muskit} measure test-suite thoroughness. All of these adapt classical testing techniques to quantum programs. None exploits the \emph{algebraic structure of the quantum system itself} as a test oracle. For VQE circuits, the properties of parametrised rotation gates and diagonal Hamiltonians impose deterministic constraints at known parameter values, constraints that are checkable in $O(1)$ circuit evaluations and require no ground-truth reference, yet this resource remains untapped.

We present \tool{}, a metamorphic testing framework that closes this gap by deriving test oracles directly from the quantum mechanical properties of VQE circuits. The key insight is that, at $\theta = 0$, every rotation gate reduces to the identity, and the quantum state is fully determined by the fixed structure of the circuit ~\cite{nielsen2010quantum}. In this regime, several quantities take exactly known values: the energy, the x-magnetisation, and the gradient with respect to each parameter. From these properties, we derive five invariants rooted in the algebra of rotation gates~\cite{schuld2019evaluating} and the structure of diagonal Hamiltonians~\cite{farhi2014qaoa,wiersema2020exploring}. Each invariant holds for any correct circuit under exact (noiseless) statevector simulation; a violation therefore guarantees a bug, giving a 0\% false-positive rate under exact, noiseless simulation. On real hardware, shot noise and gate errors may require noise-aware tolerances (Section~\ref{sec:discussion}). Because all checks run at step~0, they are unaffected by convergence failures and cost only $O(p)$ circuit evaluations ($p$ = number of parameters). The scope of \tool{} is VQE circuits with diagonal (ZZ-type) Hamiltonians and single-qubit ansatze, the dominant pattern in combinatorial-optimisation VQE; extension to molecular Hamiltonians and entangling ansatze is left to future work. To evaluate these invariants, we follow the mutation testing methodology of Fortunato et al.~\cite{fortunato2022mutation}: we inject controlled faults into correct VQE circuits and measure the \emph{mutation score}, the fraction of non-equivalent mutants detected.

We address three research questions: \textbf{RQ1}: Can physics-derived invariants detect VQE-specific faults without a ground-truth oracle, and which invariants target which fault classes? \textbf{RQ2}: Do physics-based invariants provide better diagnostic discriminability (Youden's $J$) than convergence-based testing? \textbf{RQ3}: Are the invariants sound (zero false positives) on a diverse benchmark of unmodified circuits, and what is their computational overhead?

Our contributions are:
\begin{enumerate}
\item We derive and prove five physics-based invariants for VQE circuits, each based on a different property of rotation gates and diagonal Hamiltonians. We show when each invariant applies (QAOA vs.\ hardware-efficient ansatz) and prove that the false-positive rate is 0\% under exact, noiseless statevector simulation.
\item We evaluate \tool{} via mutation testing on 500 AgentQ~\cite{jern2025agentq} circuits with five VQE-specific mutation operators: 56.9\% mutation score at 0\% FPR (Youden's $J = 0.57$), compared to $J = 0.02$ for convergence. Combining both reaches 83.9\%.
\item On all 500 unmodified benchmark circuits, L1--L5 (oracle-free) and L6 (spectral check, requires $E_\text{exact}$) each produce zero flags, confirming the 0\% false-positive rate in practice.
\end{enumerate}

The rest of this paper is organised as follows. Section~\ref{sec:related} covers related work. Section~\ref{sec:background} provides background. Section~\ref{sec:approach} presents the five invariants with proofs. Sections~\ref{sec:eval} describes the experimental setup and results. Section~\ref{sec:discussion} discusses implications and limitations, and Section~\ref{sec:conclusion} concludes.

\section{Related Work}\label{sec:related}

\begin{table*}[t]
\caption{Comparison of quantum software testing approaches. ``Oracle-free'' means no ground-truth output is needed. ``Deterministic'' means the verdict is exact (not statistical). ``VQE-specific'' means the method targets variational circuit structure.}
\label{tab:related}
\centering
\footnotesize
\setlength{\tabcolsep}{4.5pt}
\begin{tabular}{l l l c c c c}
\toprule
\textbf{Approach} & \textbf{Technique} & \textbf{Oracle source} & \textbf{Oracle-free} & \textbf{Determ.} & \textbf{VQE-spec.} & \textbf{Cost} \\
\midrule
QuiTO / Ali et al.~\cite{wang2021quito,ali2021assessing}   & Coverage-guided testing    & Expected output         & No  & Yes & No  & $O(n_\text{tests})$ \\
QDiff~\cite{wang2022qdiff}            & Differential testing       & Cross-platform agree.   & Yes & Yes & No  & $O(n_\text{back.})$ \\
Huang \& Mart.~\cite{huang2019statistical} & Statistical assertions & Distribution patterns   & Yes & No  & No  & $O(n_\text{shots})$ \\
Honarvar et al.~\cite{honarvar2020property} & Property-based testing & Universal properties    & Yes & Yes & No  & $O(n_\text{props})$ \\
Fortunato; Muskit~\cite{fortunato2022mutation,mendiluze2022muskit} & Mutation testing/analysis & Developer-written tests & No  & Yes & No  & $O(n_\text{mut.})$ \\
\midrule
\textbf{\tool{} (ours)}               & \textbf{Physics-based invariants} & \textbf{Quantum mechanics} & \textbf{Yes} & \textbf{Yes} & \textbf{Yes} & $\mathbf{O(p)}$ \\
\bottomrule
\end{tabular}
\end{table*}

Table~\ref{tab:related} positions \tool{} among existing quantum software testing methods. We organise the literature along three dimensions that define our research gap: (1)~how existing methods handle the oracle problem, (2)~whether metamorphic or property-based approaches have been applied to VQE, and (3)~how mutation testing has been used in the quantum domain.

\subsection{Quantum Software Testing and the Oracle Gap}

Quantum software engineering~\cite{piattini2020talavera,zhao2021quantum} has produced several testing techniques, each addressing the oracle problem differently. Wang et al.~\cite{wang2021quito} (QuiTO) and Ali et al.~\cite{ali2021assessing} use coverage-guided generation but require expected outputs, leaving the oracle problem unsolved. QDiff~\cite{wang2022qdiff} avoids explicit oracles via cross-platform differential testing, but depends on multiple backends producing identical results. Huang and Martonosi~\cite{huang2019statistical} propose oracle-free statistical assertions, but these are inherently probabilistic: they require many measurement shots and provide confidence bounds rather than deterministic verdicts. Paltenghi and Pradel~\cite{paltenghi2023bugs4q} catalogue real quantum bugs (Bugs4Q) but do not propose detection methods. Miranskyy et al.~\cite{miranskyy2021testing} identify the oracle problem as the central barrier.

The gap is clear: no existing method provides \emph{deterministic}, \emph{oracle-free} verdicts that exploit the \emph{physics of the specific quantum algorithm} under test. All current approaches either need expected outputs, require multiple backends, or give probabilistic guarantees. \tool{} fills this gap for VQE.

\subsection{Metamorphic and Property-Based Testing for Quantum Programs}

In the quantum domain, Honarvar et al.~\cite{honarvar2020property} apply property-based testing to Q\# programs, checking universal properties like reversibility and unitarity. These checks apply to \emph{any} quantum program but are too generic to catch VQE-specific faults such as gate substitutions or parameter corruption in variational circuits. Our approach differs in a key way: rather than checking universal quantum properties, we derive relations from the specific algebraic structure of VQE circuits, namely the commutation relations of rotation gates and diagonal Hamiltonians, and the deterministic behaviour at known parameter values. This yields invariants that are both \emph{domain-specific} (targeting VQE) and \emph{deterministic} (exact under noiseless simulation).

\subsection{Mutation Testing for Quantum Software}

Fortunato et al.~\cite{fortunato2022mutation} bring mutation testing to the quantum domain with Qiskit-level operators (gate substitution, gate removal). Mendiluze et al.~\cite{mendiluze2022muskit} extend this with Muskit, supporting multiple platforms. Both evaluate developer-written test suites, not physics-derived invariants. We adopt their methodology, using mutation testing as our \emph{evaluation framework}, but the object of evaluation is different: we assess physics-based invariants rather than manually written tests.

\subsection{LLM-Generated Quantum Circuits}

Lemieux et al.~\cite{lemieux2023codamosa} and Deng et al.~\cite{deng2023llm_testing} show that LLMs can assist with classical software testing. In the quantum domain, Jern et al.~\cite{jern2025agentq} introduce AgentQ, which fine-tunes LLMs to generate VQE circuits for combinatorial optimisation. As LLM-generated circuits enter production pipelines, automated validation that does not depend on manual test oracles becomes essential. Our work addresses this need: we validate LLM-generated circuits using physics-based invariants.

\section{Background}\label{sec:background}

This section defines the technical concepts that our invariants build on: VQE circuits and ansatz families, the oracle problem, metamorphic testing, and mutation testing.

\subsection{Variational Quantum Eigensolver}

The VQE~\cite{peruzzo2014vqe,mcclean2016theory} combines a quantum processor with a classical optimiser to approximate the ground-state energy of a Hamiltonian $H$ acting on $n$ qubits. It prepares a parametrised state $|\psi(\theta)\rangle = U(\theta)|0\rangle^{\otimes n}$ and evaluates:
\begin{equation}\label{eq:vqe}
E(\theta) = \langle 0|^{\otimes n} U^\dagger(\theta)\,H\,U(\theta)|0\rangle^{\otimes n}
\end{equation}
A classical optimiser iteratively adjusts $\theta$ to reduce $E(\theta)$. The variational principle~\cite{mcclean2016theory} guarantees $E(\theta) \geq E_0$ for every $\theta$, where $E_0$ is the exact ground-state energy.

The parametrised unitary $U(\theta)$ is called the \emph{ansatz}. Two common ansatz families appear in practice:
\begin{itemize}
\item \textbf{QAOA}~\cite{farhi2014qaoa}: initialises all qubits in $|+\rangle$ via Hadamard gates and applies layers of parametrised $R_z$ rotations. At $\theta = 0$, all rotations reduce to the identity, leaving qubits in $|+\rangle^{\otimes n}$.
\item \textbf{Hardware-efficient ansatz (HEA)}~\cite{kandala2017hardware}: uses $R_x$ and $R_z$ (or sometimes $R_y$) rotations without Hadamard initialisation. At $\theta = 0$, qubits remain in $|0\rangle^{\otimes n}$.
\end{itemize}

Both families use parametrised rotation gates $R_\alpha(\theta) = e^{-i\theta P_\alpha/2}$ for $\alpha \in \{x,y,z\}$, where $P_\alpha$ is the corresponding Pauli operator~\cite{nielsen2010quantum}. The algebraic properties of these gates, their behaviour under complex conjugation, their generators, and how they commute with diagonal Hamiltonians~\cite{schuld2019evaluating,mitarai2018quantum,stokes2020quantum}, are the foundation of our invariants.

\subsection{The Oracle Problem and Convergence}

The ground-state energy $E_0$ is both the quantity the VQE computes and the reference a test would need, creating a circular dependency~\cite{miranskyy2021testing}. Classical diagonalisation can provide $E_0$ for small instances but costs $O(2^n)$, making it infeasible at scale. The standard fallback, convergence-based testing, checks whether the VQE energy reaches the expected value within a fixed budget. This is unreliable because barren plateaus~\cite{mcclean2018barren,cerezo2021barren} flatten the cost landscape exponentially with qubit count, and local minima trap the optimiser regardless of budget~\cite{grant2019initialization}.

\subsection{Metamorphic Testing}

Metamorphic testing~\cite{chen1998metamorphic,chen2018metamorphic,segura2016survey} addresses the oracle problem by checking whether \emph{pairs} of outputs satisfy a known relationship rather than verifying individual outputs against a reference. A metamorphic relation (MR) specifies how a transformation of the input should transform the output; any violation implies a fault.

A textbook example is $\sin(-x) = -\sin(x)$: one can verify this without knowing $\sin(x)$. Our invariants apply this principle to VQE circuits. Strictly, L2 is a classical MR (comparing $E(\theta)$ and $E(-\theta)$), while L3--L5 are \emph{fixed-point assertions} at the distinguished input $\theta = 0$. Segura et al.~\cite{segura2016survey} classify identity-input checks as a special case of MRs. L1 (structural inspection) falls outside this taxonomy; we include it for completeness but do not claim it as a metamorphic relation.

\subsection{Mutation Testing}

Mutation testing~\cite{demillo1978hints,jia2011mutation} evaluates test-suite quality by introducing small, deliberate faults into the program and checking whether the suite detects each one. A detected mutant is \emph{killed}; the \emph{mutation score} (fraction killed) measures fault-detection completeness. This rests on the competent programmer hypothesis~\cite{demillo1978hints}: real bugs resemble small perturbations of correct code, so a suite that catches small perturbations is likely to catch real bugs.

\section{The \tool{} Invariant Suite}\label{sec:approach}

The premise of \tool{} is straightforward: the laws of quantum mechanics place deterministic constraints on any correctly constructed circuit, and those constraints can be repurposed as test oracles. At known parameter values, a correct VQE circuit must yield specific, analytically predictable outputs. \tool{} verifies a suite of such constraints at step~0, before the VQE optimisation loop begins. Each invariant is rooted in the algebra of rotation gates and the structure of diagonal (ZZ-type) Hamiltonians common in combinatorial optimisation. Because every constraint is a direct consequence of quantum mechanics and applies to all correct circuits, any violation signals a genuine fault, ensuring a 0\% false-positive rate under exact, noiseless simulation.

\begin{figure}[t]
\centering
\includegraphics[width=\columnwidth]{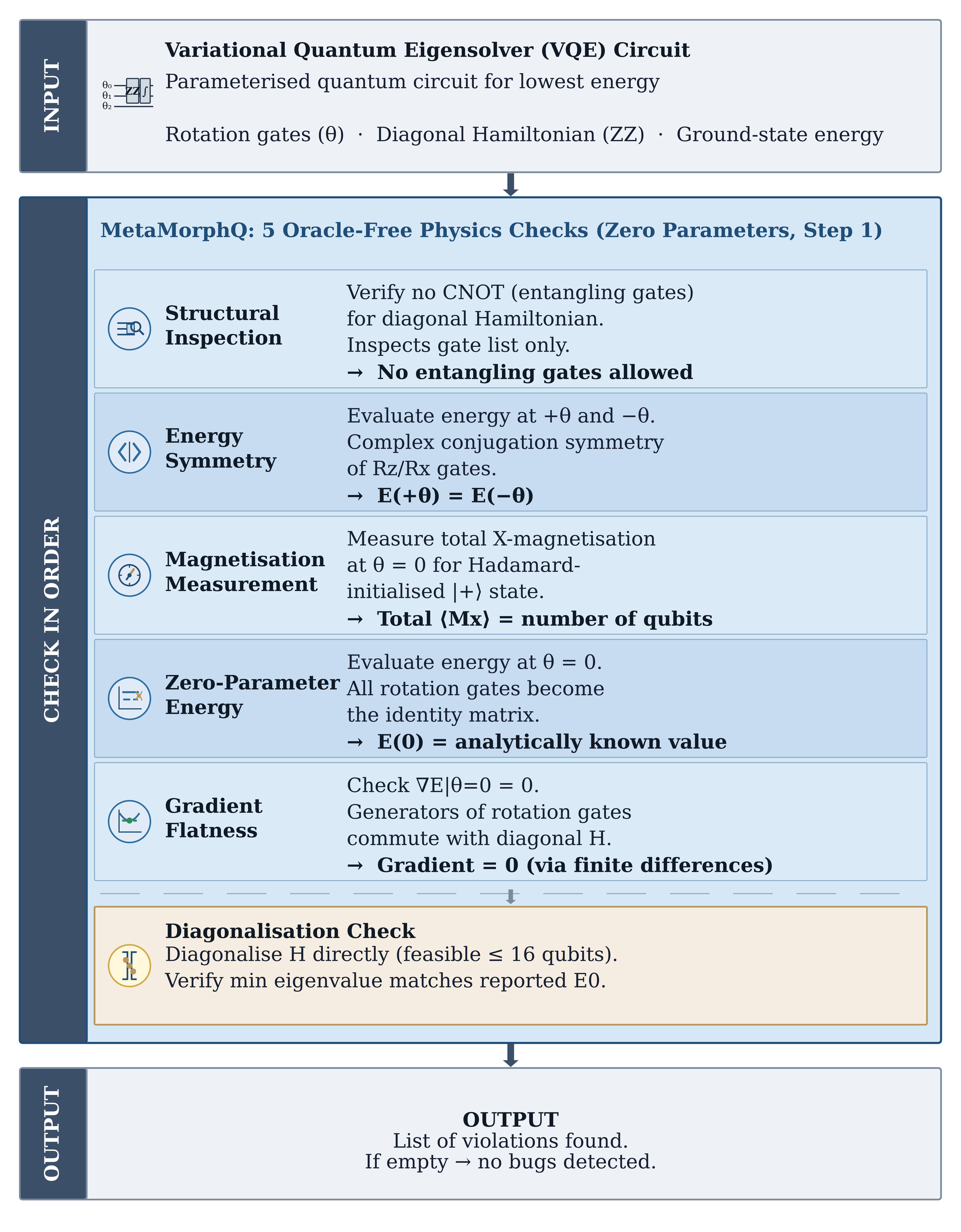}
\caption{\tool{} checking pipeline. L1--L5 are oracle-free physics invariants evaluated at step~0; L6 is a classical preflight check that requires $E_\text{exact}$.}
\label{fig:overview}
\end{figure}

Table~\ref{tab:invariants} summarises the five oracle-free invariants (L1--L5) and one classical preflight check (L6). Fig.~\ref{fig:overview} illustrates the checking pipeline: the input (circuit gates, Hamiltonian, qubit count) passes through L1--L5 at step~0 with no oracle; L6 runs as a classical preflight if $E_\text{exact}$ is available. A violation at any layer constitutes a bug report. We now describe each invariant with its formal derivation and applicability scope.

\begin{table}[t]
\caption{The \tool{} invariant suite. $p$ = number of parametrised gates. $E_\mathrm{exp}$ denotes the analytically expected energy at $\theta=0$.}
\label{tab:invariants}
\centering
\footnotesize
\setlength{\tabcolsep}{2.5pt}
\begin{tabular}{l l l r l}
\toprule
\textbf{ID} & \textbf{Property} & \textbf{Scope} & \textbf{Ev.} & \textbf{Detects} \\
\midrule
L1 & CX structure & All & 0 & extra gate \\
L2 & Parity $E(\!-\!\theta) \!=\! E(\theta)$ & $R_z$/$R_x$ & 2 & gate type, param \\
L3 & X-mag $\sum\!\langle X_i\rangle \!=\! n$ & QAOA & 1 & gate type, param \\
L4 & $E(0) \!=\! E_\mathrm{exp}$ & All & 1 & qubit idx, extra \\
L5 & $\partial E/\partial\theta_k|_0 \!=\! 0$ & Diag.\ $H$ & $2p$ & gate type, param \\
\midrule
L6 & $\lambda_\mathrm{min}(H) \!\approx\! E_\mathrm{exact}$ & $n \!\leq\! 16$ & 0$^\dagger$ & ham sign \\
\bottomrule
\multicolumn{5}{l}{\footnotesize $^\dagger$Classical diagonalisation; requires $E_\mathrm{exact}$.}
\end{tabular}
\end{table}

\subsection{L1: CX Structure Check (0 evaluations)}

For diagonal (ZZ) Hamiltonians $H = \sum_{ij} J_{ij} Z_i Z_j$ at small qubit counts, variational ansatze commonly consist of single-qubit rotations only, with the multi-qubit Hamiltonian structure handled through the expectation-value measurement~\cite{farhi2014qaoa,jern2025agentq}. In such circuits, any CNOT gate is spurious and indicates a structural error. The derivation is given in Appendix~\ref{app:L1}.

This check needs zero circuit evaluations and catches mutations that add spurious entangling gates.

\subsection{L2: Energy Parity (2 evaluations)}

For circuits containing only $R_z$ and $R_x$ parametrised gates with a real-valued Hamiltonian, the energy function exhibits a parity symmetry:
\begin{equation}\label{eq:parity}
E(-\theta) = E(\theta) \quad \forall\theta
\end{equation}
We check this relation at $\theta = \pm\pi/4 \cdot \mathbf{1}$. Circuits containing $R_y$ gates are excluded from this check. The derivation, based on the complex-conjugation properties of $R_z$ and $R_x$ gates~\cite{nielsen2010quantum,schuld2019evaluating}, is given in Appendix~\ref{app:L2}.

\subsection{L3: X-Magnetisation (1 evaluation)}

For QAOA circuits with Hadamard initialisation, the total x-magnetisation at $\theta = 0$ must equal the number of qubits:
\begin{equation}\label{eq:xmag}
\textstyle\sum_{i=0}^{n-1} \langle X_i \rangle \big|_{\theta=0} = n
\end{equation}

The intuition is that at $\theta = 0$ all rotations become identity, leaving qubits in $|+\rangle$, the $+1$ eigenstate of $X$. We check this by evaluating the auxiliary observable $H_X = \sum_i X_i$ at zero parameters (one circuit evaluation). The derivation is in Appendix~\ref{app:L3}.

\subsection{L4: Zero-Parameter Identity (1 evaluation)}

At $\theta = 0$, the circuit reduces to its fixed structure, and the energy takes an analytically known value:
\begin{equation}\label{eq:zeroparam}
E(0) = \begin{cases} 0 & \text{(QAOA)} \\ \sum_i c_i & \text{(HEA)} \end{cases}
\end{equation}
where $c_i$ are the coefficients of the diagonal Hamiltonian $H = \sum_i c_i P_i$ (each $P_i$ a product of $Z$ and $I$ operators). The derivation follows from the Pauli eigenvalue equations $\langle +|Z|+\rangle = 0$ and $\langle 0|Z|0\rangle = 1$; for Hamiltonians with non-diagonal terms ($X$ or $Y$), the expected value must be computed from the specific Pauli decomposition. See Appendix~\ref{app:L4}.

\subsection{L5: Zero-Gradient (2 evaluations per gate)}

For diagonal Hamiltonians with $R_z$ parametrised gates, or $R_x$ gates in a Hadamard sandwich ($H \cdot R_x(\theta) \cdot H$), the gradient of the energy with respect to every parameter vanishes at $\theta = 0$:
\begin{equation}\label{eq:zerograd}
\frac{\partial E}{\partial \theta_k}\bigg|_{\theta=0} = 0 \quad \forall k
\end{equation}
We evaluate this via finite differences at $\pm\pi/4$: $[E(\pi/4 \cdot \mathbf{e}_k) - E(-\pi/4 \cdot \mathbf{e}_k)] / (\pi/2)$. By the parity symmetry of L2, $E(\theta) = E(-\theta)$ for correct circuits, so this difference is zero. A gate substitution ($R_z \to R_y$) that breaks parity also produces a non-zero finite difference, which L5 detects independently of L2 on a per-parameter basis. This requires two circuit evaluations per parametrised gate. The derivation is in Appendix~\ref{app:L5}.

\subsection{L6: Hamiltonian Spectral Check (classical)}

For instances with $n \leq 16$ qubits, we compute $\lambda_\text{min}(H)$ via sparse diagonalisation and check that it matches the reported ground-state energy:
\[
|\lambda_\text{min}(H) - E_\text{exact}| \leq 0.01
\]
Unlike L1--L5, this is not a metamorphic relation: it needs a reference value ($E_\text{exact}$) and runs classically. We include it as a preflight check on the problem specification itself, catching cases where the Hamiltonian or reported energy is wrong.

\section{Evaluation}\label{sec:eval}
We evaluate \tool{} using mutation testing following Fortunato et al.~\cite{fortunato2022mutation}. This section describes the experimental setup and then answers the three research questions.

\subsection{Mutation Operators}

We define five mutation operators modelling realistic VQE bugs~\cite{demillo1978hints}. Four target the circuit: \texttt{gate\_type} swaps a rotation gate ($R_z \leftrightarrow R_y$) or initialisation gate ($H \leftrightarrow X$); \texttt{qubit\_index} moves a gate to a different wire; \texttt{extra\_gate} inserts a spurious CNOT; and \texttt{param\_corrupt} adds $\pi/4$ to one rotation angle. The fifth, \texttt{ham\_sign}, targets the Hamiltonian by negating one coefficient. Each operator is applied once per instance with a fixed seed for reproducibility.

\subsection{Dataset}

We use 500 instances from AgentQ~\cite{jern2025agentq}, a dataset of LLM-generated VQE circuits for combinatorial optimisation. The instances span 2--16 qubits across 12 problem types (hypermaxcut, min\_cut, vertex\_cover, etc.) with QAOA ($R_z$-only: 33\%, $R_y$-only: 25\%) and HEA ($R_x$/$R_z$: 28\%, $R_y$: 10\%, mixed: 3\%) ansatze. All circuits use single-qubit gates only ($n_\text{CX} = 0$), which is natural for diagonal Hamiltonians at small qubit counts~\cite{farhi2014qaoa}. This means L1 reduces to checking for the absence of any CNOT; generalisation to entangling circuits would require a different structural invariant (see Section~\ref{sec:discussion}). Because the benchmark is entirely LLM-generated, evaluation on human-written or transpiled VQE circuits remains an open item, which we revisit as a threat to validity in Section~\ref{sec:discussion}.

\subsection{Protocol}

For each of the 500 instances, we run six VQE executions: one correct and five mutated (one per operator). Each mutant applies exactly one operator; compound faults are not tested, following standard mutation testing practice~\cite{jia2011mutation}. Each execution uses Adam with learning rate 0.1 and a 400-step budget, in PennyLane~\cite{bergholm2018pennylane} with the \texttt{lightning.qubit} backend. The invariants are checked once at step~0.

A mutant is \emph{killed} if any invariant flags it. Mutants whose modification has no observable effect on the energy landscape are \emph{equivalent} and excluded~\cite{jia2011mutation}: specifically, a mutant is equivalent when the mutation leaves the Hamiltonian matrix unchanged (e.g., negating a zero coefficient) or when the mutated circuit computes the same energy function as the original. This leaves 2,469 non-equivalent mutants across 3,000 runs.

\subsection{Metrics}

We report three metrics:
\begin{itemize}
\item \emph{Mutation score}: fraction of non-equivalent mutants killed. Higher is better.
\item \emph{False-positive rate (FPR)}: fraction of correct circuits wrongly flagged. Lower is better.
\item \emph{Unique kills}: mutants killed by exactly one invariant, measuring non-redundancy.
\end{itemize}
To compare overall quality, we use Youden's $J = \text{mutation score} - \text{FPR}$~\cite{youden1950index}. $J = 1$ is perfect; $J = 0$ is no better than random guessing.

\subsection{Results}

\subsubsection{RQ1: What Is the Mutation Score?}

\begin{figure}[t]
\centering
\includegraphics[width=\columnwidth]{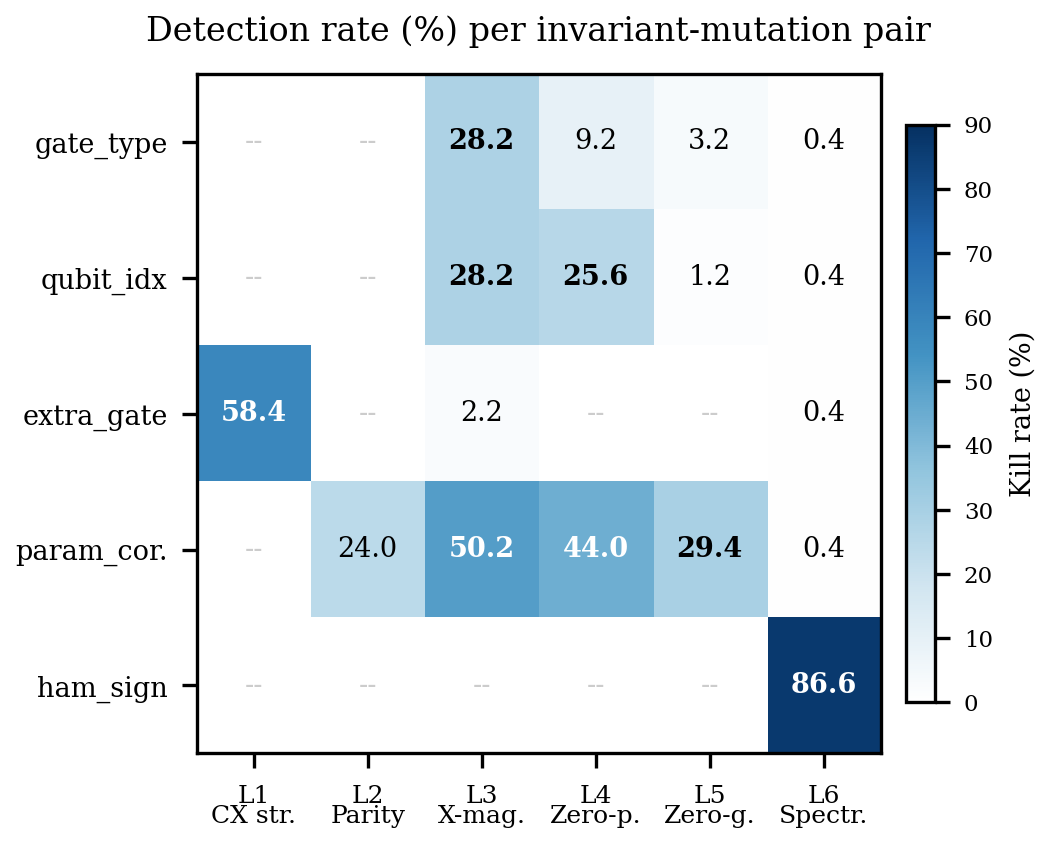}
\caption{Kill rate (\%) per invariant--mutation pair. Each invariant targets a distinct fault class, with minimal overlap.}
\label{fig:heatmap}
\end{figure}

Overall, the suite kills 56.9\% of 2,469 non-equivalent mutants with 0\% false positives on 500 correct circuits. Fig.~\ref{fig:heatmap} shows the kill rate for each invariant--mutation pair. The invariants show complementary detection patterns. L1 (CX structure) is the only check that catches extra-gate mutations (58.4\%), since it looks at gate structure rather than energy values. L3 (x-magnetisation) is the most broadly useful circuit check, catching gate-type (28.2\%), qubit-index (28.2\%), and parameter corruption (50.2\%). L6 (spectral check) is the only one that catches Hamiltonian faults (86.6\%), since circuit-level checks cannot see problem-encoding errors. L2 (energy parity) and L5 (zero-gradient) mainly target parameter corruption (24.0\% and 29.4\%).

Every invariant contributes unique kills (mutants caught by that invariant alone): L6 accounts for 408, L1 for 281, L3 for 121, L4 for 115, L5 for 21, and L2 for 5. In total, 67.7\% of all kills are unique to one invariant, meaning removing any single check would lower the overall score. This per-invariant breakdown doubles as an ablation: the distinct, largely non-overlapping kill contributions in Fig.~\ref{fig:heatmap} show that each of L1--L6 is necessary, and that dropping any one would reduce the suite's overall fault coverage.

\subsubsection{RQ2: How Does \tool{} Compare to Convergence Testing?}

We compare \tool{} against convergence-based testing, the standard approach for VQE. Convergence flags a circuit as buggy if $|E_\text{final} - E_\text{exact}| > 0.05$ after optimisation. Table~\ref{tab:baseline} compares the two.

\begin{table}[t]
\caption{Comparison of \tool{} vs.\ convergence-based testing on 2,469 non-equivalent mutants and 500 correct circuits. $J$ = Youden's J (mutation score $-$ FPR).}
\label{tab:baseline}
\centering
\footnotesize
\setlength{\tabcolsep}{3pt}
\begin{tabular}{l rrr rrr}
\toprule
& \multicolumn{3}{c}{\textbf{Convergence}} & \multicolumn{3}{c}{\textbf{\tool{}}} \\
\cmidrule(lr){2-4} \cmidrule(lr){5-7}
\textbf{Mutation} & Kill & FPR & $J$ & Kill & FPR & $J$ \\
\midrule
\texttt{gate\_type}    & 62.6 & 58.8 & 0.04 & 31.8 & 0.0 & 0.32 \\
\texttt{qubit\_idx}    & 64.2 & 58.8 & 0.05 & 28.6 & 0.0 & 0.29 \\
\texttt{extra\_gate}   & 65.2 & 58.8 & 0.06 & 58.8 & 0.0 & 0.59 \\
\texttt{param\_cor.}   & 63.4 & 58.8 & 0.05 & 80.6 & 0.0 & 0.81 \\
\texttt{ham\_sign}     & 49.3 & 58.8 & $-$0.10 & 86.6 & 0.0 & 0.87 \\
\midrule
\textbf{Overall}       & 61.1 & 58.8 & 0.02 & 56.9 & 0.0 & 0.57 \\
\textbf{Combined}      & \multicolumn{6}{c}{83.9\% mutation score} \\
\bottomrule
\end{tabular}
\end{table}

Convergence has a slightly higher raw mutation score (61.1\% vs.\ 56.9\%), but at the cost of a 58.8\% false-positive rate, flagging 294 of 500 correct circuits as buggy. Its $J = 0.02$ means it is barely better than random guessing. For Hamiltonian faults (\texttt{ham\_sign}), convergence is \emph{worse} than random ($J = -0.10$), because sign-flipped Hamiltonians often have reachable local minima. A concrete instance illustrates the gap: for one min\_cut instance, negating a single Hamiltonian coefficient yields a spectrum whose minimum the optimiser still reaches within tolerance, so convergence reports the mutant as correct; L6, by contrast, detects that $\lambda_\text{min}(H)$ no longer matches $E_\text{exact}$ and flags the fault immediately. Such encoding faults are invisible to convergence but caught deterministically by \tool{}.

The 58.8\% FPR is with a 400-step budget. More steps would help, but barren plateaus~\cite{mcclean2018barren} cause exponentially vanishing gradients that no finite budget fully resolves. \tool{}'s invariants run once at step~0 and are unaffected by the budget.

\begin{figure}[t]
\centering
\includegraphics[width=\columnwidth]{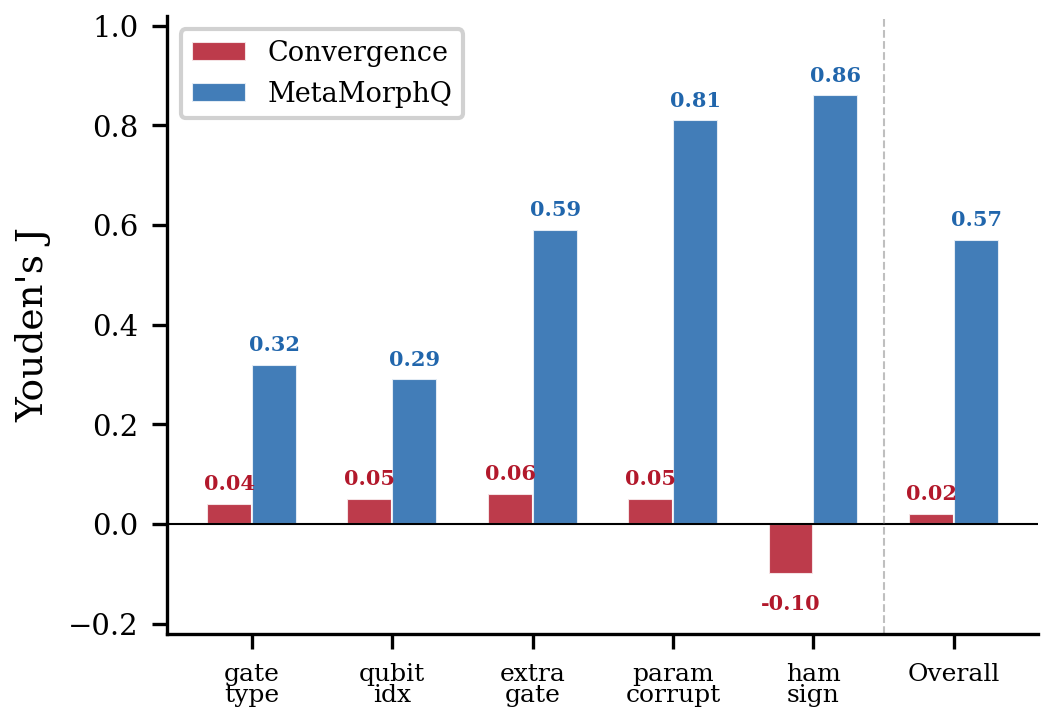}
\caption{Youden's $J$ per mutation type. \tool{} (blue) provides a strong diagnostic signal across all fault classes, while convergence (red) hovers near zero and is negative for Hamiltonian faults.}
\label{fig:youdens}
\end{figure}

\tool{} achieves $J = 0.57$ versus $J = 0.02$ for convergence ($\Delta J = 0.55$). For parameter corruption ($J = 0.81$) and Hamiltonian sign flips ($J = 0.87$), the signal is strong. Fig.~\ref{fig:youdens} visualises this contrast across all mutation types.

The two methods complement each other well. A mutant counts as killed by the combined approach if \emph{either} method flags it (union of kill sets). Convergence uniquely kills 667 mutants (faults visible only during optimisation), while \tool{} uniquely kills 564 (faults visible at step~0 but masked by convergence noise). Together they reach 83.9\%, a 22.8 and 27.0 percentage-point gain over either alone. This shows they target different fault classes and should be used together.

\subsubsection{RQ3: Soundness and Cost}

To validate soundness beyond mutation testing, we run the full suite on all 500 unmodified circuits. L1--L5 (oracle-free invariants) produce zero flags, confirming the 0\% FPR predicted by the proofs in Section~\ref{sec:approach}. L6 (spectral check) also produces zero flags: for every instance, $|\lambda_{\min}(H) - E_\mathrm{exact}| < 10^{-6}$, confirming that the AgentQ dataset's reported ground-state energies are consistent with the Hamiltonians.

Across all 500 circuits and all six invariants, no correct circuit is flagged, giving a 0\% FPR in practice. This matches the theoretical guarantee: each invariant follows from quantum mechanical laws that hold for any correct circuit under exact, noiseless simulation.

For cost, L1 and L6 require no circuit execution; L2 uses 2 evaluations (at $\theta = \pm\pi/4$); L3 and L4 share a single evaluation at $\theta = 0$; and L5 uses $2p$ evaluations via finite differences. The total is $3 + 2p$ circuit evaluations at step~0 ($p$ = number of parametrised gates). For the median AgentQ circuit ($p = 8$), this is 19 evaluations vs.\ $\sim$400 for convergence, roughly 20$\times$ cheaper. The oracle-free invariants L1--L5 need no ground-truth value, while convergence requires $E_\text{exact}$.

\section{Discussion}\label{sec:discussion}
The results show that physics-based invariants provide a reliable, low-cost testing signal where convergence-based testing falls short. In this section, we interpret the key findings, examine the detection limits, and discuss the broader implications and threats to validity.

\subsection{Why Youden's J, Not Raw Mutation Score}

Comparing raw mutation scores (61.1\% vs.\ 56.9\%) would suggest convergence is better. However, this ignores false positives. A test that labels \emph{every} circuit as buggy gets 100\% mutation score while being useless. Youden's $J$ corrects for this by subtracting the FPR.

With $J = 0.02$, convergence gives almost no useful information: a ``failure'' verdict gives a developer little confidence that the circuit is actually buggy. \tool{}'s $J = 0.57$ means a flag is very likely a real bug. A test that cannot tell correct circuits from faulty ones is not useful, no matter how many faults it catches.

\subsection{Surviving Mutants and Detection Limits}

The 43.1\% surviving mutants are mostly \emph{step-0 equivalent}: invisible when parameters are zero. For example, $R_z(0) = R_y(0) = I$~\cite{nielsen2010quantum}, so swapping them is undetectable at $\theta = 0$. Similarly, qubit-index swaps on symmetric Hamiltonians leave $E(0)$ unchanged. These can only be caught by running the full optimisation.

This is a trade-off. \tool{} gives up recall on faults that only show during optimisation, in exchange for deterministic, $O(1)$-cost, zero-false-positive detection at step~0. The 83.9\% combined score shows this trade-off works: \tool{} catches what convergence misses (Hamiltonian faults, parameter corruptions), convergence catches what \tool{} misses (step-0 equivalent substitutions), and together they cover most faults.

\subsection{Physics as a Source of Test Oracles}

A key contribution of this work is showing that the algebraic structure of quantum circuits can serve directly as a testing resource. Each invariant exploits a different property: conjugation symmetry of rotation gates (L2), Pauli eigenvalue structure (L3, L4), commutativity of diagonal operators (L5), and gate-level structural constraints (L1). These are not ad-hoc checks but consequences of quantum mechanics, which is why they guarantee zero false positives under exact simulation.

To our knowledge, this \emph{deterministic} physics-as-oracle approach is new in quantum software testing. Prior work either adapts classical techniques (coverage, property-based testing) to quantum programs, or uses statistical checks on measurement distributions~\cite{huang2019statistical}. Our results show that going the other direction, deriving oracles from the mathematical properties of the system under test, is productive and complementary. The specific invariants here target VQE with diagonal Hamiltonians, but the methodology applies to any quantum algorithm where correctness implies verifiable constraints.

\subsection{Extending the Invariant Suite}

The present suite is deliberately scoped to diagonal Hamiltonians and single-qubit ansatze, but the underlying methodology suggests several further invariants. For molecular Hamiltonians with $XX$ and $YY$ terms, the $\theta=0$ state is still analytically known, so an L4-style identity could be derived from the full Pauli decomposition rather than the diagonal special case. For entangling ansatze, the absence-of-CNOT check (L1) would generalise to a CNOT-parity or commuting-block check that verifies the entangling structure matches the intended $R_{zz}$ decomposition. Symmetry-based invariants are another candidate: many physical Hamiltonians possess particle-number or $Z_2$ symmetries that a correct ansatz must preserve, giving conserved quantities checkable at arbitrary parameter values rather than only at $\theta=0$. Finally, multi-point invariants evaluated at several distinguished parameter settings (e.g., $\theta = \pi$) could extend coverage to faults that are step-0 equivalent. We leave the formal derivation and evaluation of these variants to future work.

\subsection{Practical Impact}

As LLM-generated quantum circuits become more common~\cite{jern2025agentq,lemieux2023codamosa,deng2023llm_testing}, fast automated validation becomes important. \tool{}'s invariants need $O(1)$ evaluations and no oracle, making them a good fit for CI/CD pipelines in quantum software. A developer can run the suite at step~0 before committing to a full VQE optimisation, catching structural, parametric, and encoding faults immediately with zero false positives. The 0\% false-positive rate on all 500 unmodified circuits means that every flag is actionable: a developer who sees a violation knows it is a genuine bug, not a convergence artefact.

\subsection{Threats to Validity}

\emph{Internal validity.} Our evaluation uses diagonal (ZZ) Hamiltonians with single-qubit ansatze. This covers most QAOA and simple HEA circuits for combinatorial optimisation, but not all VQE applications. L2 and L5 rely on commutation relations specific to diagonal Hamiltonians; extending them to molecular Hamiltonians with $XX$ or $YY$ terms would need new derivations. L1 would shift from checking absence of CNOTs to checking CNOT parity in circuits that legitimately use entangling gates. L3, L4, and L6 generalise directly to any Hamiltonian, since they depend only on the $\theta = 0$ state and eigenvalue structure. The 0\% FPR guarantee holds under exact (noiseless) simulation; on real quantum hardware, shot noise and gate errors could cause small deviations that cross tolerance thresholds, requiring noise-aware tolerances.

\emph{External validity.} The benchmark is entirely LLM-generated (AgentQ). These circuits may share systematic biases from the fine-tuning process that are not representative of hand-written VQE implementations. To mitigate this, the dataset spans 12 problem types, two ansatz families, and 2--16 qubits. Nonetheless, generalisability to human-authored and transpiled circuits is not established by our experiments; evaluating \tool{} on hand-written and compiler-transpiled VQE benchmarks is a priority for future work, and we expect the physics-derived invariants to transfer because they depend on circuit semantics rather than on how the circuit was produced.

\emph{Construct validity.} We exclude mutants with no observable effect, but some surviving mutants may be equivalent in a subtler sense (identical energy landscapes despite different gate sequences). Standard practice~\cite{jia2011mutation} is to accept this as a known limitation of mutation testing. Each mutant applies a single operator; compound faults are not tested. The convergence comparison uses a 400-step budget with Adam; a different optimiser or budget would change the convergence FPR, so the reported baseline is specific to this configuration, though barren plateaus~\cite{mcclean2018barren} are a structural problem that no finite budget fully resolves. The physics-based invariants are unaffected by these choices.

\section{Conclusion}\label{sec:conclusion}

We presented \tool{}, a testing framework that derives oracle-free test verdicts from the algebraic properties of VQE circuits. Using the structure of rotation gates and diagonal Hamiltonians, we defined five invariants, each rooted in a different quantum mechanical principle, that can be checked in $O(1)$ evaluations at step~0 with no ground-truth reference.

On 500 AgentQ circuits with 2,469 non-equivalent mutants, the suite achieves 56.9\% mutation score at 0\% false positives ($J = 0.57$), compared to $J = 0.02$ for convergence. The two methods complement each other, reaching 83.9\% combined. On all 500 unmodified circuits, the full suite produces zero flags, confirming the 0\% false-positive rate in practice. We frame this contribution deliberately: \tool{} is a physics-based testing method for VQE circuits with diagonal Hamiltonians and single-qubit ansatze, the dominant pattern in combinatorial-optimisation VQE, rather than a general-purpose framework for every VQE circuit.

Within that scope, our results show that the mathematical structure of quantum circuits provides a useful, largely untapped source of test oracles. Future work will extend the approach to non-diagonal Hamiltonians and entangling circuits, investigate multi-step invariants at non-zero parameter values, evaluate \tool{} on human-written and transpiled benchmarks, and integrate \tool{} into CI/CD pipelines for quantum software.

\section*{Acknowledgment}
 
This work was partially supported by the Australian Research Council (ARC) through Linkage Project LP240100523, \emph{Blockchain Based Quantum Safe for Secure Digital Medical Passport}. The authors gratefully acknowledge the support of the project partners and collaborators.

\bibliographystyle{IEEEtran}
\bibliography{nhi_qsw_26_refs}

@article{peruzzo2014vqe,
  author    = {A. Peruzzo and J. McClean and P. Shadbolt and M.-H. Yung and X.-Q. Zhou and P. J. Love and A. Aspuru-Guzik and J. L. O'Brien},
  title     = {A Variational Eigenvalue Solver on a Photonic Quantum Processor},
  journal   = {Nature Communications},
  volume    = {5},
  pages     = {4213},
  year      = {2014},
  doi       = {10.1038/ncomms5213}
}

@article{farhi2014qaoa,
  author    = {E. Farhi and J. Goldstone and S. Gutmann},
  title     = {A Quantum Approximate Optimization Algorithm},
  journal   = {arXiv preprint arXiv:1411.4028},
  year      = {2014}
}

@article{kandala2017hardware,
  author    = {A. Kandala and A. Mezzacapo and K. Temme and M. Takita and M. Brink and J. M. Chow and J. M. Gambetta},
  title     = {Hardware-Efficient Variational Quantum Eigensolver for Small Molecules and Quantum Magnets},
  journal   = {Nature},
  volume    = {549},
  number    = {7671},
  pages     = {242--246},
  year      = {2017},
  doi       = {10.1038/nature23879}
}

@article{tilly2022vqe_review,
  author    = {J. Tilly and H. Chen and S. Cao and D. Picozzi and K. Setia and Y. Li and E. Grant and L. Wossnig and I. Rungger and G. H. Booth and J. Tennyson},
  title     = {The Variational Quantum Eigensolver: A Review of Methods and Best Practices},
  journal   = {Physics Reports},
  volume    = {986},
  pages     = {1--128},
  year      = {2022},
  doi       = {10.1016/j.physrep.2022.08.003}
}

@article{mcclean2016theory,
  author    = {J. R. McClean and J. Romero and R. Babbush and A. Aspuru-Guzik},
  title     = {The Theory of Variational Hybrid Quantum-Classical Algorithms},
  journal   = {New Journal of Physics},
  volume    = {18},
  number    = {2},
  pages     = {023023},
  year      = {2016},
  doi       = {10.1088/1367-2630/18/2/023023}
}

@article{preskill2018nisq,
  author    = {J. Preskill},
  title     = {Quantum Computing in the {NISQ} Era and Beyond},
  journal   = {Quantum},
  volume    = {2},
  pages     = {79},
  year      = {2018},
  doi       = {10.22331/q-2018-08-06-79}
}

@article{cerezo2021barren,
  author    = {M. Cerezo and A. Sone and T. Volkoff and L. Cincio and P. J. Coles},
  title     = {Cost Function Dependent Barren Plateaus in Shallow Parametrized Quantum Circuits},
  journal   = {Nature Communications},
  volume    = {12},
  pages     = {1791},
  year      = {2021},
  doi       = {10.1038/s41467-021-21728-w}
}

@article{mcclean2018barren,
  author    = {J. R. McClean and S. Boixo and V. N. Smelyanskiy and R. Babbush and H. Neven},
  title     = {Barren Plateaus in Quantum Neural Network Training Landscapes},
  journal   = {Nature Communications},
  volume    = {9},
  number    = {1},
  pages     = {4812},
  year      = {2018},
  doi       = {10.1038/s41467-018-07090-4}
}

@book{nielsen2010quantum,
  author    = {M. A. Nielsen and I. L. Chuang},
  title     = {Quantum Computation and Quantum Information},
  edition   = {10th Anniversary},
  publisher = {Cambridge University Press},
  year      = {2010},
  doi       = {10.1017/CBO9780511976667}
}

@article{schuld2019evaluating,
  author    = {M. Schuld and V. Bergholm and C. Gogolin and J. Izaac and N. Killoran},
  title     = {Evaluating Analytic Gradients on Quantum Hardware},
  journal   = {Physical Review A},
  volume    = {99},
  number    = {3},
  pages     = {032331},
  year      = {2019},
  doi       = {10.1103/PhysRevA.99.032331}
}

@article{mitarai2018quantum,
  author    = {K. Mitarai and M. Negoro and M. Kitagawa and K. Fujii},
  title     = {Quantum Circuit Learning},
  journal   = {Physical Review A},
  volume    = {98},
  number    = {3},
  pages     = {032309},
  year      = {2018},
  doi       = {10.1103/PhysRevA.98.032309}
}

@article{stokes2020quantum,
  author    = {J. Stokes and J. Izaac and N. Killoran and G. Carleo},
  title     = {Quantum Natural Gradient},
  journal   = {Quantum},
  volume    = {4},
  pages     = {269},
  year      = {2020},
  doi       = {10.22331/q-2020-05-25-269}
}

@article{wiersema2020exploring,
  author    = {R. Wiersema and C. Zhou and Y. de Sereville and J. F. Carrasquilla and Y. B. Kim and H. Yuen},
  title     = {Exploring Entanglement and Optimization within the {Hamiltonian} Variational Ansatz},
  journal   = {PRX Quantum},
  volume    = {1},
  number    = {2},
  pages     = {020319},
  year      = {2020},
  doi       = {10.1103/PRXQuantum.1.020319}
}

@article{grant2019initialization,
  author    = {E. Grant and L. Wossnig and M. Ostaszewski and M. Benedetti},
  title     = {An Initialization Strategy for Addressing Barren Plateaus in Parametrized Quantum Circuits},
  journal   = {Quantum},
  volume    = {3},
  pages     = {214},
  year      = {2019},
  doi       = {10.22331/q-2019-12-09-214}
}

@article{chen1998metamorphic,
  author    = {T. Y. Chen and S. C. Cheung and S. M. Yiu},
  title     = {Metamorphic Testing: A New Approach for Generating Next Test Cases},
  journal   = {Technical Report HKUST-CS98-01, Department of Computer Science, HKUST},
  year      = {1998}
}

@article{chen2018metamorphic,
  author    = {T. Y. Chen and F.-C. Kuo and H. Liu and P.-L. Poon and D. Towey and T. H. Tse and Z. Q. Zhou},
  title     = {Metamorphic Testing: A Review of Challenges and Opportunities},
  journal   = {ACM Computing Surveys},
  volume    = {51},
  number    = {1},
  pages     = {1--27},
  year      = {2018},
  doi       = {10.1145/3143561}
}

@article{segura2016survey,
  author    = {S. Segura and G. Fraser and A. B. Sanchez and A. Ruiz-Cort\'{e}s},
  title     = {A Survey on Metamorphic Testing},
  journal   = {IEEE Transactions on Software Engineering},
  volume    = {42},
  number    = {9},
  pages     = {805--824},
  year      = {2016},
  doi       = {10.1109/TSE.2016.2532875}
}

@article{jia2011mutation,
  author    = {Y. Jia and M. Harman},
  title     = {An Analysis and Survey of the Development of Mutation Testing},
  journal   = {IEEE Transactions on Software Engineering},
  volume    = {37},
  number    = {5},
  pages     = {649--678},
  year      = {2011},
  doi       = {10.1109/TSE.2010.62}
}

@article{demillo1978hints,
  author    = {R. A. DeMillo and R. J. Lipton and F. G. Sayward},
  title     = {Hints on Test Data Selection: Help for the Practicing Programmer},
  journal   = {Computer},
  volume    = {11},
  number    = {4},
  pages     = {34--41},
  year      = {1978},
  doi       = {10.1109/C-M.1978.218136}
}

@inproceedings{wang2022qdiff,
  author    = {X. Wang and P. Arcaini and T. Yue and S. Ali},
  title     = {{QDiff}: Differential Testing of Quantum Software Stacks},
  booktitle = {Proceedings of the 37th IEEE/ACM International Conference on Automated Software Engineering (ASE)},
  pages     = {1--12},
  year      = {2022},
  doi       = {10.1145/3551349.3556967}
}

@inproceedings{wang2021quito,
  author    = {X. Wang and P. Arcaini and T. Yue and S. Ali},
  title     = {{QuiTO}: A Coverage-Guided Test Generation Tool for Quantum Programs},
  booktitle = {Proceedings of the 36th IEEE/ACM International Conference on Automated Software Engineering (ASE)},
  pages     = {1237--1241},
  year      = {2021},
  doi       = {10.1109/ASE51524.2021.9678798}
}

@inproceedings{fortunato2022mutation,
  author    = {D. Fortunato and J. Campos and R. Abreu},
  title     = {Mutation Testing of Quantum Programs Written in {QISKit}},
  booktitle = {Proceedings of the IEEE/ACM 44th International Conference on Software Engineering: New Ideas and Emerging Results (ICSE-NIER)},
  pages     = {1--5},
  year      = {2022},
  doi       = {10.1145/3510455.3512791}
}

@inproceedings{honarvar2020property,
  author    = {S. Honarvar and N. K. Mousavi and R. Nagarajan},
  title     = {Property-Based Testing of Quantum Programs in {Q\#}},
  booktitle = {Proceedings of the IEEE/ACM 42nd International Conference on Software Engineering: New Ideas and Emerging Results (ICSE-NIER)},
  pages     = {430--435},
  year      = {2020},
  doi       = {10.1145/3377816.3381731}
}

@inproceedings{paltenghi2023bugs4q,
  author    = {M. Paltenghi and M. Pradel},
  title     = {{Bugs4Q}: A Benchmark of Real Bugs for Quantum Programs},
  booktitle = {Proceedings of the 45th International Conference on Software Engineering (ICSE)},
  pages     = {1--12},
  year      = {2023},
  doi       = {10.1109/ICSE48619.2023.00199}
}

@inproceedings{ali2021assessing,
  author    = {S. Ali and P. Arcaini and X. Wang and T. Yue},
  title     = {Assessing the Effectiveness of Input and Output Coverage Criteria for Testing Quantum Programs},
  booktitle = {Proceedings of the 14th IEEE Conference on Software Testing, Verification and Validation (ICST)},
  pages     = {13--23},
  year      = {2021},
  doi       = {10.1109/ICST49551.2021.00014}
}

@inproceedings{miranskyy2021testing,
  author    = {A. Miranskyy and L. Zhang and J. Rilling},
  title     = {On Testing Quantum Programs},
  booktitle = {Proceedings of the IEEE/ACM 43rd International Conference on Software Engineering: New Ideas and Emerging Results (ICSE-NIER)},
  pages     = {56--60},
  year      = {2021},
  doi       = {10.1109/ICSE-NIER52604.2021.00021}
}

@inproceedings{huang2019statistical,
  author    = {Y. Huang and M. Martonosi},
  title     = {Statistical Assertions for Validating Patterns and Finding Bugs in Quantum Programs},
  booktitle = {Proceedings of the 46th International Symposium on Computer Architecture (ISCA)},
  pages     = {541--553},
  year      = {2019},
  doi       = {10.1145/3307650.3322213}
}

@article{zhao2021quantum,
  author    = {J. Zhao},
  title     = {Quantum Software Engineering: Landscapes and Horizons},
  journal   = {arXiv preprint arXiv:2007.07047},
  year      = {2021}
}

@article{piattini2020talavera,
  author    = {M. Piattini and G. Peterssen and R. P\'{e}rez-Castillo and J. L. Hevia and M. A. Serrano and G. Hern\'{a}ndez and others},
  title     = {The {Talavera Manifesto} for Quantum Software Engineering and Programming},
  journal   = {ACM SIGSOFT Software Engineering Notes},
  volume    = {45},
  number    = {4},
  pages     = {1--5},
  year      = {2020},
  doi       = {10.1145/3417564.3417567}
}

@inproceedings{lemieux2023codamosa,
  author    = {C. Lemieux and J. P. Inala and S. K. Lahiri and S. Sen},
  title     = {{CodaMosa}: Escaping Coverage Plateaus in Test Generation with Pre-Trained Large Language Models},
  booktitle = {Proceedings of the 45th International Conference on Software Engineering (ICSE)},
  pages     = {919--931},
  year      = {2023},
  doi       = {10.1109/ICSE48619.2023.00085}
}

@inproceedings{deng2023llm_testing,
  author    = {Y. Deng and C. S. Xia and H. Peng and C. Yang and L. Zhang},
  title     = {Large Language Models Are Zero-Shot Fuzzers: Fuzzing Deep-Learning Libraries via Large Language Models},
  booktitle = {Proceedings of the 32nd ACM SIGSOFT International Symposium on Software Testing and Analysis (ISSTA)},
  pages     = {423--435},
  year      = {2023},
  doi       = {10.1145/3597926.3598067}
}

@article{bergholm2018pennylane,
  author    = {V. Bergholm and J. Izaac and M. Schuld and C. Gogolin and S. Ahmed and V. Ajith and M. S. Alam and G. Alonso-Linaje and B. AkashNarayanan and A. Asber and others},
  title     = {{PennyLane}: Automatic Differentiation of Hybrid Quantum-Classical Computations},
  journal   = {arXiv preprint arXiv:1811.04968},
  year      = {2018}
}

@inproceedings{jern2025agentq,
   title={Agent-Q: Fine-Tuning Large Language Models for Quantum Circuit Generation and Optimization},
   DOI={10.1109/qce65121.2025.00179},
   booktitle={2025 IEEE International Conference on Quantum Computing and Engineering (QCE)},
   author={Jern, Linus and Uotila, Valter and Yu, Cong and Zhao, Bo},
   year={2025}
}

@article{youden1950index,
  author    = {W. J. Youden},
  title     = {Index for Rating Diagnostic Tests},
  journal   = {Cancer},
  volume    = {3},
  number    = {1},
  pages     = {32--35},
  year      = {1950},
  doi       = {10.1002/1097-0142(1950)3:1<32::AID-CNCR2820030106>3.0.CO;2-3}
}

@inproceedings{mendiluze2022muskit,
  author    = {E. Mendiluze and S. Ali and P. Arcaini and T. Yue},
  title     = {{Muskit}: A Mutation Analysis Tool for Quantum Software Testing},
  booktitle = {Proceedings of the IEEE International Conference on Quantum Computing and Engineering (QCE)},
  pages     = {1--12},
  year      = {2022},
  doi       = {10.1109/QCE53715.2022.00042}
}

\appendices

\section*{Note on the Appendix}
Each derivation below assembles well-known identities from quantum mechanics and the research literature into the specific form used by our test oracles. We provide them in full for reproducibility and to make the source of each step explicit.

\section{L1: CX Structure Check}\label{app:L1}

\emph{Sources.} The structure of QAOA ansatze for diagonal Hamiltonians is given in Farhi et al.~\cite{farhi2014qaoa}, Section~II. The commutativity of diagonal operators is covered in Nielsen \& Chuang~\cite{nielsen2010quantum}, Section~2.1.9.

\emph{Derivation.} For a diagonal Hamiltonian $H = \sum_{ij} J_{ij} Z_i Z_j$, the cost unitary $e^{-i\gamma H}$ factorises into commuting terms $\prod_{ij} e^{-i\gamma J_{ij} Z_i Z_j}$ because all $Z_i Z_j$ are simultaneously diagonal. In general, each two-qubit factor $e^{-i\gamma J_{ij} Z_i Z_j}$ requires a CNOT-based decomposition. However, at small qubit counts the standard approach in LLM-generated circuits~\cite{jern2025agentq} is to use a variational ansatz of single-qubit rotations ($R_z$, $R_x$) only, with the multi-qubit Hamiltonian structure handled entirely through the expectation-value measurement rather than explicit entangling gates. In the AgentQ dataset, all 500 circuits follow this pattern, giving $n_\text{CX} = 0$. For such circuits, any CNOT gate is spurious and indicates a structural fault.

Note: this check is specific to single-qubit ansatze. For circuits that legitimately use entangling gates (e.g., CNOT-$R_z$-CNOT decompositions of $R_{zz}$), L1 would need to verify CNOT parity rather than absence (see Section~\ref{sec:discussion}). $\square$

\section{L2: Energy Parity}\label{app:L2}

\emph{Sources.} The complex-conjugation properties of rotation gates ($R_z(-\theta) = R_z(\theta)^*$, $R_x(-\theta) = R_x(\theta)^*$) follow from the matrix exponential definitions in Nielsen \& Chuang~\cite{nielsen2010quantum}, Section~4.2 (``Single qubit operations''), using the fact that $Z$ and $X$ are real matrices while $Y$ is purely imaginary. The application of this conjugation symmetry to parameter-shift gradients appears in Schuld et al.~\cite{schuld2019evaluating}, Section~II.

\emph{Derivation.} From the matrix exponential $R_z(\theta) = e^{-i\theta Z/2}$, entry-wise complex conjugation gives $R_z(\theta)^* = e^{+i\theta Z/2} = R_z(-\theta)$, since $Z$ is real. The same holds for $R_x$. All non-parametric gates in the circuit (Hadamard, Pauli-$X$) are also real-valued matrices, so the full unitary satisfies $U(-\theta) = U(\theta)^*$. For a real Hamiltonian $H$:
\[
E(-\theta) = \langle\psi(\theta)^* | H | \psi(\theta)^*\rangle = E(\theta)^* = E(\theta)
\]
where the last equality holds because the energy is real. This symmetry breaks for $R_y$ gates: since $Y$ is purely imaginary, $R_y(\theta)$ is a real matrix, giving $R_y(\theta)^* = R_y(\theta)$ and therefore $R_y(-\theta) = R_y(\theta)^\top \neq R_y(\theta)^*$. Substituting $R_z \to R_y$ thus violates the conjugation property, breaking parity and enabling fault detection. $\square$

\section{L3: X-Magnetisation}\label{app:L3}

\emph{Sources.} The identity $R_\alpha(0) = I$ is immediate from the matrix exponential definition (Nielsen \& Chuang~\cite{nielsen2010quantum}, Section~4.2). The Pauli eigenvalue equation $X|+\rangle = |+\rangle$ (equivalently, $\langle +|X|+\rangle = 1$) is a standard result found in Nielsen \& Chuang~\cite{nielsen2010quantum}, Section~2.1.3 (``The Pauli matrices'').

\emph{Derivation.} At $\theta = 0$, every parametrised rotation gate reduces to the identity: $R_\alpha(0) = e^{-i \cdot 0 \cdot P_\alpha/2} = I$. For QAOA circuits, the Hadamard gate maps $|0\rangle \to |+\rangle$, and $|+\rangle$ is the $+1$ eigenstate of $X$, so $\langle +|X|+\rangle = 1$ for each qubit. Summing over all $n$ qubits gives $\sum_i \langle X_i \rangle = n$.

Any modification to the initialisation layer disrupts this value. For example, replacing $H \to X$ gives $X|0\rangle = |1\rangle$, and $\langle 1|X|1\rangle = 0$, producing a clear deviation from $n$. $\square$

\section{L4: Zero-Parameter Identity}\label{app:L4}

\emph{Sources.} The expectation values $\langle +|Z|+\rangle = 0$ and $\langle 0|Z|0\rangle = 1$ are Pauli eigenvalue equations from Nielsen \& Chuang~\cite{nielsen2010quantum}, Section~2.1.3. The application to QAOA energy evaluation at $\theta = 0$ is described in Farhi et al.~\cite{farhi2014qaoa}, Section~II, where the authors note that the initial state $|+\rangle^{\otimes n}$ has zero expectation value for every $Z_iZ_j$ term.

\emph{Derivation.} For QAOA, qubits are initialised in $|+\rangle^{\otimes n}$. Since $|+\rangle = (|0\rangle + |1\rangle)/\sqrt{2}$ is an equal superposition of $Z$-eigenstates, $\langle +|Z_i|+\rangle = 0$, and therefore every interaction term satisfies $\langle Z_iZ_j \rangle = \langle Z_i\rangle\langle Z_j\rangle = 0$ (the qubits are in a product state). This gives $E(0) = \sum_{ij} J_{ij} \cdot 0 = 0$.

For HEA with a diagonal Hamiltonian, qubits remain in $|0\rangle^{\otimes n}$ at $\theta = 0$. Since $|0\rangle$ is the $+1$ eigenstate of $Z$, we have $\langle 0|Z_i Z_j|0\rangle = (+1)(+1) = 1$ for all pairs, giving $E(0) = \sum_i c_i$.

(For Hamiltonians containing non-diagonal terms such as $X_i$ or $Y_i$, the expectation values at $\theta = 0$ would differ and must be computed from the specific Pauli decomposition of $H$.)

Any structural error, such as a misplaced gate or a spurious entangling operation, alters the $\theta = 0$ state and produces a different energy. $\square$

\section{L5: Zero-Gradient}\label{app:L5}

\emph{Sources.} The parameter-shift rule ($\partial E/\partial \theta_k = \frac{1}{2}[E(\theta_k + \pi/2) - E(\theta_k - \pi/2)]$) was independently derived by Schuld et al.~\cite{schuld2019evaluating} (Section~II, Theorem~1) and Mitarai et al.~\cite{mitarai2018quantum} (Section~III). The commutativity of diagonal operators ($[Z_iZ_j, Z_k] = 0$) is standard (Nielsen \& Chuang~\cite{nielsen2010quantum}, Section~2.1.9). The Hadamard-basis transformation $HXH = Z$ is given in Nielsen \& Chuang~\cite{nielsen2010quantum}, Section~4.2.

\emph{Derivation.} We use the finite-difference approximation at step size $\epsilon = \pi/4$. At $\theta = 0$, we evaluate $E(\pi/4 \cdot \mathbf{e}_k)$ and $E(-\pi/4 \cdot \mathbf{e}_k)$, where $\mathbf{e}_k$ is the $k$-th unit vector. The zero-gradient property follows from two independent arguments:

\emph{Via parity (L2).} For circuits with $R_z$ and $R_x$ gates only, L2 establishes $E(-\theta) = E(\theta)$. Applying this with $\theta = \pi/4 \cdot \mathbf{e}_k$ gives $E(\pi/4 \cdot \mathbf{e}_k) = E(-\pi/4 \cdot \mathbf{e}_k)$, so the finite difference vanishes.

\emph{Via commutators (independent argument).} Since all parametric gates are identity at $\theta = 0$, the state before gate $k$ is determined solely by the fixed initialisation layer ($U_\text{pre}$, e.g.\ Hadamards). The exact gradient is proportional to $\langle \psi_0 | U_\text{pre}^\dagger [H, G_k] U_\text{pre} | \psi_0 \rangle$, where $G_k$ is the generator of gate $k$. For $R_z$ gates with $G_k = Z_k$ and a diagonal $H = \sum_{ij} J_{ij} Z_i Z_j$: all operators are diagonal, so $[H, Z_k] = 0$ identically, giving zero gradient regardless of $U_\text{pre}$. For $R_x$ gates in QAOA circuits: $U_\text{pre}$ consists of Hadamards, and the $R_x$ gate appears in a Hadamard sandwich ($H \cdot R_x(\theta) \cdot H$), so the effective generator transforms as $H X_k H = Z_k$~\cite{nielsen2010quantum}, which again commutes with $H$. \emph{Precondition:} the $R_x$ Hadamard-sandwich structure must hold; an $R_x$ gate outside this structure would have generator $X_k$ with $[Z_iZ_j, X_k] \neq 0$.

\emph{Detection.} Substituting $R_z \to R_y$ breaks both arguments: (1)~parity fails (see L2), so the finite-difference test yields a non-zero value; (2)~the generator $Y_k$ satisfies $[Z_iZ_j, Y_k] \neq 0$ when $k \in \{i, j\}$. While L5 and L2 share the parity mechanism, L5 localises the fault to a specific parameter $k$, providing per-gate diagnostic information that L2 cannot. $\square$

\end{document}